# Building a scalable python distribution for HEP data analysis

David J Lange

Physics Department, Princeton University, Princeton, NH 08544, USA

E-mail: dlange@princeton.edu

**Abstract**. There are numerous approaches to building analysis applications across the high-energy physics community. Among them are Python-based, or at least Python-driven, analysis workflows. We aim to ease the adoption of a Python-based analysis toolkit by making it easier for non-expert users to gain access to Python tools for scientific analysis. Experimental software distributions and individual user analysis have quite different requirements. Distributions tend to worry most about stability, usability and reproducibility, while the users usually strive to be fast and nimble. We discuss how we built and now maintain a python distribution for analysis while satisfying requirements both a large software distribution (in our case, that of CMSSW) and user, or laptop, level analysis. We pursued the integration of tools used by the broader data science community as well as HEP developed (e.g., histogrammar, root_numpy) Python packages. We discuss concepts we investigated for package integration and testing, as well as issues we encountered through this process. Distribution and platform support are important topics. We discuss our approach and progress towards a sustainable infrastructure for supporting this Python stack for the CMS user community and for the broader HEP user community.

## 1. Introduction

Use of the Python programming language has become ubiquitous in high-energy physics (HEP) and more broadly across data science. Python, as well as its user community, has evolved significantly since it was introduced into experiment software frameworks more than a decade ago. For example, the initial motivations to introduce Python as a dependency of the CMS software (CMSSW) distribution [1] were for runtime job configuration management, and for analysis using a light-weight version of the CMS framework (FWLite) [2] and/or the PyROOT toolkit [3].

These proceedings describe an effort to build a modern and maintainable Python toolkit matching the needs of both HEP frameworks and HEP analysts. The primary project goals include

1. **Reduce the startup burden for HEP analysts** interested in applying a new Python analysis tool to their project. If up to date Python-based tools are available in the same experimental software distributions that are used to access common HEP tools such as ROOT as well as experimental specific software, then the time needed for analysts to gain access to these tools can be considerably reduced.

2. **Facilitate use of a standard software environment** within analysis groups or experimental collaborations. HEP working groups typically consist of a geographically distributed community, typically using local computing resources for day-to-day work. The work of these groups can be simplified if they use a standard software platform, such as that typically maintained by central experimental teams. This approach reduces the chance to obtain different, or even incorrect, results due to inconsistent software versions or installations. Another aspect is the interoperability of software versions. Instead of small analysis groups each needing to understand the requirements and interdependencies of Python toolkits, a common software stack can be maintained and verified by an expert before put into use.

3. **Usability of distributed computing systems**. There is an increasing number of reasons why running Python-based toolkits on "the Grid" or more generically on a non-local set of computing resources is interesting. This will continue as the suite of computationally efficient data science tools available continues to increase.

These are some of the reasons why we felt a maintained Python toolkit is needed and of interest to the HEP community. In the next sections we will describe our approach and current state of implementation. We used CMSSW and its software distribution system as the basis for our work. The concepts and much of the implementation are not specific to CMS and can be easily ported. We will discuss this approach as well as how we have made this distribution easily available to developers not using CMSSW or otherwise working outside CMS.

## 2. Our approach

The CMSSW distribution has always included distributions of packages that form a relatively complete set of dependencies in addition to the build of the experiment specific codebase itself. This ensures that the entire software stack is built in a consistent manner. It was natural to expand this distribution to include our desired set of Python tools. Our aim was to do this in a simple yet maintainable way. Our approach was to adopt standard tools and make them easier for collaborators to use, rather than to build new capabilities.

*2.1. PIP-based approach*

As the number of Python packages expanded, we quickly realized that we needed to evaluate standard solutions for integrating Python tools without having to create and maintain code for each tool. After some evaluation, we selected PIP [4] to handle the Python tool management for CMSSW. PIP is a standard tool for managing Python packages that nearly all experienced Python users are familiar with.

Using PIP and the PYPI database [5] provides CMSSW a community standard mechanism for managing Python tools. We use source distributions on PYPI whenever possible to ensure that distributed toolkit is consistent with the full software stack that CMSSW distributes. We have only a few exceptions where a source based distribution is required.

This change also allowed us to standardize and simplify the code needed each time a Python tool is to be added to CMSSW. We can specify just the package, version and its dependencies to create the new tool in CMSSW. As CMSSW is RPM based, our implementation is driven by SPEC files, which look like this:

```
### RPM external py2-numba 0.33.0
## INITENV +PATH PYTHONPATH %{i}/${PYTHON_LIB_SITE_PACKAGES}

Requires: py2-funcsigs py2-enum34 py2-six
Requires: py2-singledispatch py2-llvmlite py2-numpy

## IMPORT build-with-pip
```

The last line is a small helper function (build-with-pip) that handles all of the common functionality. This is primarily to retrieve the sources from PYPI, to check that all dependencies are satisfied, to do the actual installation (including RPM creation in the case of CMSSW), and to provide hooks to handle some special cases (i.e., when options or environment variables are needed by the installation). We chose not to use PIP itself to discover, download and install dependencies. This choice was made to retain control over the full set of packages in CMSSW to help ensure that unwanted dependencies are not added by mistake.

Other advantages include the ability to discover the available Python packages distributed by CMSSW and their versions using a standard mechanism (PIP):

```
dlange> pip list
appdirs (1.4.3)
bleach (2.0.0)
Bottleneck (1.2.1)
certifi (2017.4.17)
chardet (3.0.4)
click (6.7)
climate (0.4.6)
```

## 2.2. Allowing users to be nimble

The most notable difference between the requirements of experimental distributions and user analysts is the expected turnaround time when software changes are needed. We added the VirtualEnv toolkit [6] to make this easy to do using the same tool (PIP) as used to put together the distribution itself. This approach also makes it possible to use the already existing mechanisms in CMS for making software changes automatically distributed to analysis jobs run on the CMS distributed computing system.

We use as an example the case where a user that wants to use Theano version 0.9.0 in their analysis whereas CMSSW currently uses 0.8.2 in the version of the software the user has. First the user will check what is current available in CMSSW (using two different mechanisms):

```
dlange> pip list | grep Theano
Theano (0.8.2)

dlange> python
Python 2.7.11 (default, Apr 28 2017, 13:50:27)
>>> import theano
>>> print theano.__version__
0.8.2
```

Next, a new working area for the VirtualEnv toolkit is created, activated and its corresponding path added to the Python path.

```
dlange> virtualenv updateTheano
New python executable in /build/dlange/CMSSW_9_3_X_2017-08-07-2300/updateTheano/bin/python
Installing setuptools, pip, wheel...done.
dlange> source updateTheano/bin/activate.csh

[updateTheano] dlange> setenv PYTHONPATH $PWD/updateTheano/lib/python2.7/site-
packages/:$PYTHONPATH
```

Then PIP is used to download and install the desired new version of Theano as well as any new dependencies. The CMSSW distribution is checked first for the dependencies before they are downloaded. In this case, the existing distribution already satisfies all of the needed dependencies.

```
[updateTheano] dlange> pip install Theano==0.9.0
Collecting Theano==0.9.0
Requirement already satisfied: scipy>=0.14
Requirement already satisfied: numpy>=1.9.1
Installing collected packages: Theano
Found existing installation: Theano 0.8.2
Not uninstalling theano at /cvmfs/cms-ib.cern.ch/…....... /lib/python2.7/site-packages, outside environment
/build/dlange/CMSSW_9_3_X_2017-08-07-2300/updateTheano
Successfully installed Theano-0.9.0
```

Finally, we can confirm that the new version of Theano is properly configured for use:

```
[updateTheano] dlange> pip list | grep Theano
Theano (0.9.0)

[updateTheano] dlange> python
Python 2.7.11 (default, Apr 28 2017, 13:50:27)
>>> import theano
>>> print theano.__version__
0.9.0
```

*2.3. Installing outside of CMSSW*

Using PIP also provides a natural mechanism to distribute our set of Python tools outside of CMSSW. We created a simple PYPI package [7] that can be installed in any environment using PIP. One important use case in CMS is since few collaborators run Scientific Linux on their personal laptops, this PIP-based mechanism is one way to obtain the same Python environment locally as is provided by the CMSSW distribution. The motivation of this work was to ease the standardization of tools across users in a working group. That is, working groups could adopt this PYPI package to easily provide the baseline versions of tools shared by the working group.

**3. Python Tools**

There is a broad range of Python toolkits of potential interest for data science and HEP, including an increasing number of toolkits developed specifically for HEP analysis use cases. We have attempted to add a broad range of tools that are in use across the HEP community. We have prioritized some packages simply based on CMS user interests and requests. A full list is beyond the scope of document, but can be found in our GitHub repository [8]. Example packages that we have included can be roughly grouped into five broad areas [9]:

1. **HEP specific toolkits**: Examples include rootpy, histogrammar, root_pandas, xrootdpyfs and root_numpy
2. **Notebook support**: The Jupyter stack

3. **Data science toolkits**: Examples include Bottleneck, pandas, downhill, sympy, numba, pytables and scipy.
4. **Interfaces to data storage formats**: Examples include HDF5, AVRO and ThriftPy.
5. **Machine learning toolkits**: Examples include Keras, tensorflow, scikit-learn, theano, and theanets.

## 4. Conclusion

We have described a new Python distribution mechanism meant to satisfy needs of both HEP analysis users and HEP experimental frameworks. It is based on PIP to provide ease of maintenance as well as a standard user interface to the distribution. Our development has been performed within the CMSSW distribution but does not strongly depend on it. We continue to extend this work with the aim to keep the distribution up to date as the data science Python tool kit evolves.

The authors contribution to this work was supported by the National Science Foundation under grants ACI-1450377 and PHY-1624356.